\documentclass[aps,prl,twocolumn,groupedaddress]{revtex4}

\usepackage{graphicx}
\begin{document}

\title{Electron and Phonon Cooling in a Superconductor - Normal Metal - Superconductor Tunnel Junction}


\author{Sukumar Rajauria$^{1}$, P. S. Luo$^{1}$, T. Fournier$^{1}$, F. W. J. Hekking$^{2,3}$, H. Courtois$^{1,3}$ and B. Pannetier$^{1}$}
\affiliation{$^{1}$Institut N\' eel, CNRS and Universit\'e Joseph Fourier, 25 Avenue des Martyrs, BP 166, 38042 Grenoble, France.
$^{2}$LPMMC, Universit\'e Joseph Fourier and CNRS, 25 Avenue des Martyrs, BP 166, 38042 Grenoble, France.
$^{3}$Institut Universitaire de France.}


\date{\today}

\begin{abstract}
We present evidence for the cooling of normal metal phonons, in addition to the well-known electron cooling, by electron tunneling in a Superconductor - Normal metal - Superconductor tunnel junction. The normal metal electron temperature is extracted by comparing the device current-voltage characteristics to the theoretical prediction. We use a quantitative model for the heat transfer that includes the electron-phonon coupling in the normal metal and the Kapitza resistance between the substrate and the metal. It gives a very good fit to the data and enables us to extract an effective phonon temperature in the normal metal.
\end{abstract}


\maketitle In a Normal metal - Insulator - Superconductor (N-I-S) tunnel junction, the presence of the superconducting energy gap induces the selective tunneling of high energy quasi-particles out of the normal metal \cite{APL-Nahum,RMP-Giazotto}. This phenomenon generates a heat current from the normal metal to the superconductor that is maximum at a voltage bias just below the superconductor energy gap. Only single electron tunneling contributes to the thermal current, while Andreev reflection does not. A double tunnel junction (S-I-N-I-S) geometry optimizes the cooling power so that the electronic temperature of the normal metal can be lowered from 300 mK down to about 100 mK \cite{APL-Leivo}. This significant temperature reduction raises hope to use N-I-S junctions for on-chip cooling \cite{APL-Clark}  of nano-sized systems like high-sensitivity detectors and quantum devices.

A complete investigation of thermal effects in such hybrid nanostructures at very low temperature requires a precise determination of both the electron and phonon temperatures. A current-biased N-I-S junction on the normal metal can be used as an electron thermometer by measuring the voltage at a constant bias current \cite{APL-Chandrasekhar}. However, the phonon temperature is very difficult to measure directly and is usually assumed to be that of the substrate. As we will detail below, this
assumption hinders a full characterization of the thermal effects occurring in the device. Specifically, compared to previous studies \cite{PRL-Roukes,PRB-Wellstood}, the electrons are cooled here, which in turn can cool the phonons. The importance of this effect depends on the various energy relaxation processes, {\em i.e.} on the various thermal couplings between electrons, normal metal phonons and substrate phonons at the nanometer scale.

In this Letter, we report on a precise investigation of the current-voltage characteristics of S-I-N-I-S junctions where the coupling between the normal metal phonons and the substrate is weak. The electronic temperature is obtained directly from the current-voltage characteristic. We obtain a quantitative description of the data with a heat transfer model that includes the electron-phonon scattering in the normal metal and the Kapitza resistance between the metal and the substrate. This analysis provides an effective phonon temperature in the normal metal which until now has never been accessed in such devices.

\begin{figure}[t]
\includegraphics[width=7.5 cm]{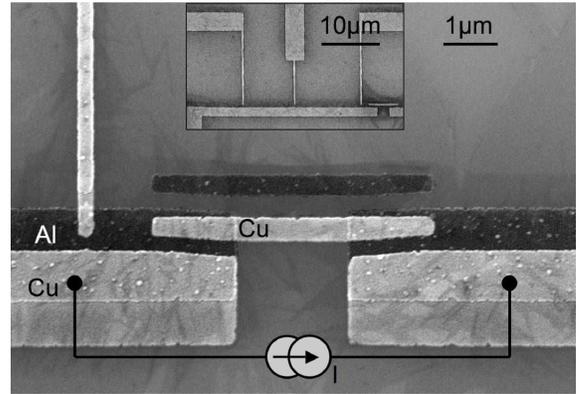}
\caption{Scanning electron micrograph of a cooler similar to Sample 1 featuring two large area Al-AlO$_{x}$-Cu junctions in series. The inset shows the complete device with 3 probes on the superconducting side of the cooler junction.} \label{Photo}
\end{figure}

In the following, we will discuss data from two typical samples with a slightly different geometry. Fig.~1 shows one of our S-I-N-I-S coolers which were fabricated using electron beam lithography, two-angle shadow evaporation and lift-off on oxidized Si wafers. The two 40 nm thick and 1.5 $\mu$m wide superconducting Al electrodes were in-situ oxidized in 0.2 mbar of O$_{2}$ for 3 min before the deposition of the $d =$ 50 nm thick, $L=$5 (Sample 1) or 4 (Sample 2) $\mu$m long and 0.3 $\mu$m wide normal metal Cu electrode. The elastic mean free path in Cu is 33 nm. The two symmetric junctions of dimensions 1.5 $\times$ 0.3 $\mu m^2$ give a total normal-state resistance in the range 2-3 k$\Omega$. The Cu electrode has been made rather short and thick. The thermal coupling between electrons and phonons is therefore increased relative to the coupling to the substrate. The overlapping extra copper shadows outside the N-I-S junctions are assumed to act as quasiparticle traps for a better thermalization of the superconducting reservoirs \cite{APL-Pekola}. In addition to the two cooling junctions, we added three Cu tunnel probes of area 0.3$\times$0.3 $\mu m^2$ on one Al electrode (see Fig. \ref{Photo} inset).

\begin{figure}[t]
\begin{center}
\includegraphics[width=8.5 cm]{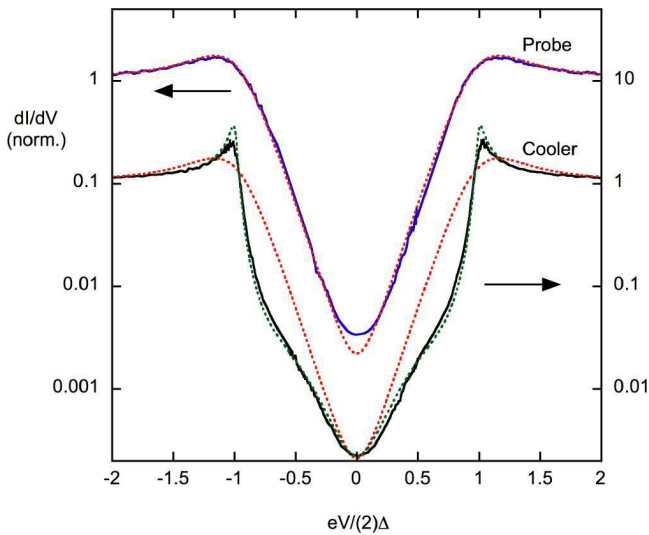}
\caption{(Color online) Differential conductance data from Sample 1 measured at a cryostat temperature of 278 mK. Top curves: data (full blue line) from the middle external probe fitted with the isotherm (dotted red line) at 300 mK with $R_{n}$ = 5.7 k$\Omega$, $\Delta$ = 0.22 meV. Bottom curves: cooler junction data (full black line) compared with the isotherm model (dotted red line) at 300 mK and the electron-phonon model (dotted green line) with $R_{n}$ = 2.8 k$\Omega$, $\Delta$ = 0.21 meV, $KA$ = 75 $pW.K^{-4}$. The voltage is normalized to $\Delta$ (top curves) and $ 2 \Delta$ (bottom curves).}
\label{CondDiff}
\end{center}
\end{figure}

We measured the current-voltage (I-V) characteristics of several S-I-N-I-S coolers in a He$^3$ cryostat. The differential conductance $dI/dV$ of every junction was obtained numerically. For sample 1, Fig.~\ref{CondDiff} shows, on a logarithmic scale, the differential conductance at a 278 mK cryostat temperature of both the middle external probe junction and the cooler double junction. The tunnel current of a N-I-S junction can be written as:
\begin{equation}
I(V)=\frac{1}{eR_{n}}\int^{\infty}_{0}n_{s}(E)[f_{N}(E-eV/2)-f_{N}(E+eV/2)]dE,
\end{equation}
where $R_{n}$ is its normal state resistance, $f_{N}$ is the electron energy distribution function in the normal metal and $n_{s}(E)=\mid E \mid /\sqrt{E^2-\Delta^2}$ is the normalized BCS density of states in the superconductor. Thus the tunnel current in a N-I-S junction does not depend on the superconductor quasiparticle distribution. The probe data is well fitted by Eq.~(1) in the hypothesis of a distribution function $f_N$ at thermal equilibrium. The fit temperature of 300 mK is only slightly higher than the measured cryostat temperature. The conductance at very small bias diverges slightly from the model while its linear slope (in log scale) versus the voltage in the sub-gap region agrees very well with the prediction. This good agreement demonstrates that the sample tunnel junctions are of good quality.

Whereas the normal metal electrode of the probe junction is strongly thermalized to the substrate temperature because of its large area, the central island of the cooler is small and thus weakly coupled to it. As can be seen in Fig.~\ref{CondDiff}, the cooler junction indeed shows a strikingly different behavior that could not be fitted by Eq.~(1) assuming thermal equilibrium at any given temperature. The experimental curve and the calculated isotherm at 300 mK coincide at zero bias, which confirms that the electron temperature is approximately equal to the cryostat temperature in the absence of current. In the subgap region, the differential conductance of the cooler is always below the isotherm. This behavior exemplifies clearly the cooling of the electronic population of the normal metal part of the device.

The phase-coherence time $\tau_{\phi}$ in Cu was measured by a weak-localization experiment in a wire fabricated with the same material. This time gives an estimate for the electron-electron scattering time. We found $\tau_{\phi}$ = 0.19 ns at 275 mK, which is much smaller than the mean escape time of an electron out of the metal island estimated to be about 100 ns. We conclude that the electrons in the normal metal island are at thermal quasi-equilibrium and that an electronic temperature $T_e$ can be well defined. We note that this regime is more easily met at the relatively high temperatures on which we concentrate here than at lower temperatures where deviations from quasi-equilibrium are expected to be prominent \cite{PRB-Heslinga}.

\begin{figure}[t]
\begin{center}
\includegraphics[width=8.5 cm]{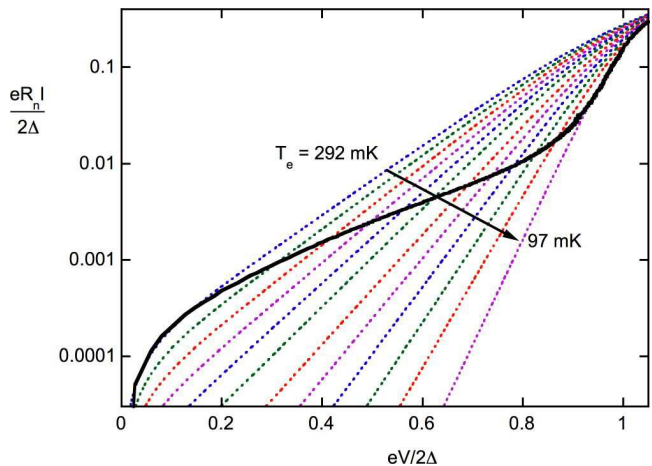}
\caption{(Color online) Sample 2 experimental current-voltage characteristic at a 275 mK cryostat temperature (full line, black) superposed on a series of calculated isotherm characteristics following Eq.~(1) (dotted lines) from T = 292 mK (top) to T = 97 mK (bottom). Every crossing point gives the electronic temperature $T_e$ in the normal metal at a particular bias.}
\label{IV}
\end{center}
\end{figure}

In the following, we assume that the two refrigerating N-I-S junctions are symmetric so that there is an equal voltage drop across each junction. Moreover, we assume the absence of leakage channels through the junction and of sub-gap energy states in the superconductor~\cite{PRL-Pekola}. The Andreev reflection contribution to the subgap current \cite{PRL-Pothier} is here negligeable. We extract the temperature $T_e(V)$ in the sub-gap region by superimposing the experimental I-V curve on a series of isotherm curves obtained from Eq.~(1), see Fig.~\ref{IV}. Every crossing point gives the electronic temperature $T_e$ in the normal metal at a particular bias. Let us note that this comparison cannot be made on the differential conductance curve since there would be a systematic error in the determination of $T_e(V)$ due to its bias dependence.

\begin{figure}[t]
\begin{center}
\includegraphics[width=8.5 cm]{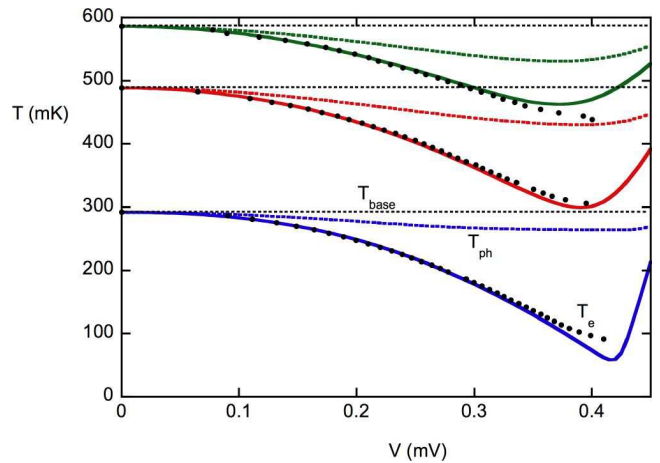}
\caption{(Color online) Sample 2 dependence of $T_e$ (dots) on the cooler bias for a cryostat temperature of 275, 470 and 570 mK. The base temperature extracted from the data is respectively 292, 489 and 586 mK. The coloured full and dotted lines give the calculated electron and phonon temperatures respectively obtained from the electron-phonon model for $\Sigma$ = 2 nW.$\mu$m$^{-3}$.K$^{-5}$, $KA$ = 66 $pW.K^{-4}$.}
\label{PhElT}
\end{center}
\end{figure}

Fig.~\ref{PhElT} displays the effective electronic temperature in the normal metal for Sample 2. Taking into account a leakage resistance of down to 5 $M \Omega$ (which is the minimum possible as based on the I-V characteristic measured at 200 mK), would not change the extracted electronic temperature $T_e(V)$ by more than 1 mK at the gap edge. As bias is increased, the electronic temperature decreases until the bias hits the gap. This minimum temperature of about 100 mK for a base temperature of about 300 mK is typical for electronic coolers probed by conventional N-I-S thermometers \cite{APL-Leivo}. At a bias above the gap, the electronic temperature cannot be accurately determined the same way since the current-voltage characteristics calculated at different temperatures overlap strongly.

\begin{figure}[t]
\begin{center}
\includegraphics[width=8.5 cm]{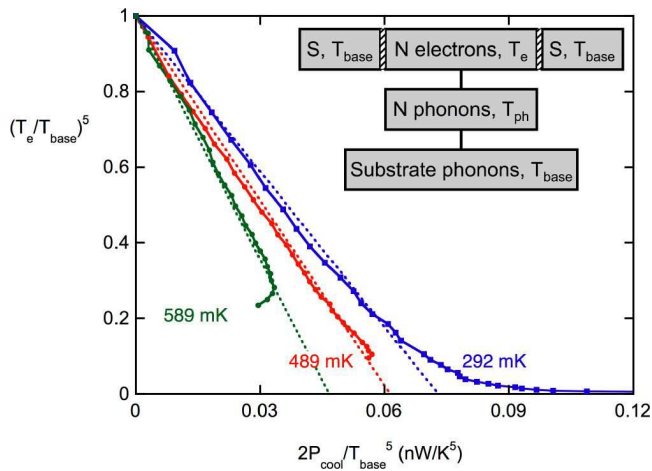}
\caption{(Color online) Sample 2 measured bias dependence of the quantity $(T_e/T_{base})^5$ for a base temperature $T_{base}$ of 292, 489 and 586 mK from bottom to top as a function of $P/T_{base}^5$. Linear fits are displayed which gives values for the $\Sigma$ parameter equal to 1.22, 1.02 and 0.78 respectively. Inset shows the schematic of the thermal model.}
\label{Power5}
\end{center}
\end{figure}

Fig.~\ref{Power5}  inset shows the schematic of the thermal model we consider, which is similar to the one described in \cite{PRL-Roukes,PRB-Wellstood} in the framework of hot electrons in metals. The normal metal electrons are at thermal equilibrium at the temperature $T_e$ and cooled by the two superconducting reservoirs thermalized at the base temperature $T_{base}$. The heat current $P_{cool}$ out of the normal metal into an individual N-I-S junction is given by:
\begin{equation}
P_{cool}=\frac{1}{e^2R_{n}}\int^{\infty}_{-\infty}(E-eV/2)n_{s}(E)[f_{N}(E-eV/2)-f_{S}(E)]dE
\end{equation}
where $f_{S}$ is the energy distribution function in the superconductor. The electron-phonon scattering is the dominant energy exchange driving energy into the electronic system with a power:
\begin{equation}
P_{e-ph}(T_{e},T_{ph})=\Sigma U(T_{e}^{5} - T_{ph}^{5})
\label{ElvsPh}
\end{equation}
where $\Sigma$ is a material dependent prefactor and $U$ is the volume of normal island. Eq. \ref{ElvsPh} is derived under the hypothesis that the phonons have an energy spectrum similar to the bulk regime and that their energy distribution is thermal at a temperature $T_{ph}$.  Its validity was recently demonstrated experimentally in Cu submicron wires \cite{JLTP-Maasilta}. According to~\cite{PRL-Perrin}, the actual distribution function will be out of equilibrium, with a lower energy fraction corresponding to the base temperature $T_{base}$ and a higher energy fraction at the electron temperature $T_e$. We verified that this out-of-equilibrium distribution function can be approximated with an equilibrium one with an effective phonon temperature $T_{ph}$ in between  $T_{base}$ and $T_{e}$. The heat balance equation for the N-metal electrons reads $2P_{cool}+P_{e-ph}=0$, where the factor 2 accounts for the presence of two symmetric N-I-S junctions. A power dissipation term due to a leakage resistance (of down to 5 $M \Omega$ which gives a power of only 2 fW) could be added to the above equation, but would not make any difference in the following discussion.

The power transfer between the phonons of the normal metal and the substrate is given by the Kapitza coupling:
\begin{equation}
P_{K}(T_{ph},T_{base})= KA(T_{ph}^{4} - T_{base}^{4}),
\end{equation}
where $A$ is the interface area and $K$ depends on the materials in contact \cite{RMP-Swartz}. Here we consider that the phonons of the substrate always remain at the base temperature $T_{base}$. The heat balance equation for the normal metal phonons is given by $P_{e-ph}+P_{K}=0$.

It is generally accepted that the phonons of a metal film are strongly thermalized to the substrate, corresponding to the limit of large $K$ in the above model, leading to $T_{ph} = T_{base}$. In this case, Eq. \ref{ElvsPh} implies that the quantity $(T_e/T_{base})^5$ depends linearly on the quantity $2P_{cool}/T_{base}^5$ with a slope equal to $1/\Sigma U$. Fig. \ref{Power5} shows that although the low bias data at a given base temperature follows reasonably such a behaviour, the related values of the parameter $\Sigma$ depend significantly on the base temperature. Moreover, they are in the range 0.8-1.2 nW.$\mu$m$^{-3}$.K$^{-5}$, i.e. well below the expected value of 2 nW.$\mu$m$^{-3}$.K$^{-5}$ \cite{PRL-Roukes,PRB-Wellstood,JLTP-Maasilta}. Thus we cannot fit our data within the thermalized phonon hypothesis. Let us point out that, in contrast with previous studies of hot electron effects, we consider here a regime with a relatively high bath temperature and electrons cooled below that temperature. 

In order to describe our experimental results, we propose to take into account a phonon cooling. We have self-consistently numerically solved the two heat balance equations as a function of sample bias in order to calculate $T_{e}(V)$ and $T_{ph}(V)$ for every base temperature. In the following, we have taken the electron-phonon coupling constant $\Sigma$ to be equal to the well-accepted value $\Sigma$ = 2 nW.$\mu$m$^{-3}$.K$^{-5}$ and used the quantity $KA$ as the fit parameter. We also used the measured values of the superconducting gap $\Delta$ and of the normal state resistance $R_{n}$. We obtained a very good fit of the bias dependence of the electron temperature at every base temperature for  $KA$ = 66 $pW.K^{-4}$ (Sample 2), see Fig.~\ref{PhElT}.  At the gap edge, it is less accurate, presumably because of mechanisms that are not included in our simple model, including the back-flow of hot quasi-particles accumulated close to the junctions. Similar results were obtained at every base temperature on all of our cooler devices with different tunnel resistances and junction areas.

Fig.~\ref{PhElT} also shows, for three different base temperatures, the calculated effective phonon temperature as a function of the sample bias. At a base temperature of 489 mK, the phonons reach at the optimum bias (with a cooling power of 1.5 pW) a temperature about 50 mK below. At lower temperature, the ratio between the phonon cooling and the electron cooling decreases because the electron-phonon decoupling ($\propto T^{-4}$) dominates the Kapitza resistance ($\propto T^{-3}$). Above the gap, the electron temperature and the phonon temperature increase, since in this regime the tunneling brings heat to the electrons. As a consistency check, we calculated the differential conductance (Eq.~(1)) of every cooler as a function of the bias, using the bias-dependent electron temperature determined above. Fig.~\ref{CondDiff} shows Sample 1 calculated differential conductance curve compared to the experimental data. The two curves show a very good agreement (note the logarithmic scale).

The parameter $KA$ accounts for the total Kapitza coupling of the Cu island to the Si substrate, that includes the contributions of both a direct Cu/Si contact and a contact through the oxidized Al film with a presumably weaker efficiency. Dividing the fit-derived $KA$ parameter by the full area of the Cu island provides an effective mean value $\langle K \rangle$. We found that $\langle K \rangle$ varies from sample to sample in the range 40 - 60 W.m$^{-2}$.K$^{-4}$. It is about a factor 3 below the values calculated for bulk Si/Cu interface \cite{RMP-Swartz} and only a factor 2 below the experimental value in Ref. \onlinecite{PRL-Roukes}.

In conclusion, the precise investigation of S-I-N-I-S coolers characteristics led us to study cold electron effects with a relatively high temperature bath, in the framework of models previously mainly used to describe hot electrons effects with a very low temperature bath. We demonstrated the relevance of the phonon cooling, which can be characterized by the dimensionless parameter $\Sigma d T /K$ here of the order of 1. The description of the relevant thermal couplings between the phonon and electron populations at the nanometer scale calls for further experimental and theoretical investigation.

The authors are grateful to Nanofab-CNRS for sample fabrication, E. Favre-Nicollin for contributing to early stages of this work and to P. Brosse for technical support. We acknowledge useful discussions with J. Pekola, P. Gandit, F. Giazotto and I. J. Maasilta. This work was funded by EU STREP project 'SFINx' and by NanoSciERA project 'Nanofridge'.

 \end{document}